\documentclass[review]{elsarticle}

\usepackage{amsmath,amssymb}
\usepackage{graphicx}
\usepackage{braket}
\usepackage{bm}
\usepackage{siunitx}
\usepackage{hyperref}
\usepackage{float}
\usepackage{booktabs}
\usepackage{xcolor}
\usepackage{listings}
\usepackage{subcaption}
\definecolor{codegray}{gray}{0.95}
\journal{Computer Physics Communications}

\begin{document}

\begin{frontmatter}

\title{\texttt{\textbf{NeutrinoOsc3Flavor}}: CP Phase Dependence in Three-Flavor Neutrino Oscillations: A Numerical Study in Vacuum and Matter }

\author[inst1]{Baktiar Wasir Farooq}
\author[inst1]{Bipin Singh Koranga}
\author[inst1]{Ansh Prasad}
\author[inst2]{Imran Khan}
\affiliation[inst1]{
  organization={Kirori Mal College, University of Delhi},
  city={Delhi},
  country={India}
}
\affiliation[inst2]{
  organization={Hindu College, University of Delhi},
  city={Delhi},
  country={India}
}

\begin{abstract}
We present NeutrinoOsc3Flavor, a lightweight and fully transparent computational framework for exact three-flavor neutrino oscillation studies in vacuum and constant-density matter. The code numerically solves the Schrödinger evolution equation in the flavor basis using explicit construction and diagonalization of the effective Hamiltonian within the PMNS formalism, including full CP-violating phase dependence.\\
In contrast to large-scale oscillation toolkits optimized for experimental simulations, NeutrinoOsc3Flavor is designed as a minimal-dependency reference implementation, emphasizing analytical traceability, equation-level accessibility, and cross-platform portability. The framework relies solely on NumPy for numerical linear algebra and runs natively on both Linux and Windows systems without external compilation or specialized libraries.\\
As an internal consistency and validation feature, we implement an independent analytical determination of the matter-modified Hamiltonian eigenvalues using the Cardano method and demonstrate excellent agreement with numerical diagonalization. CP-phase dependence is used as a sensitive diagnostic of numerical stability and correctness of the evolution operator in both vacuum and matter.\\
NeutrinoOsc3Flavor is intended as a verification-oriented and pedagogical computational tool, suitable for theoretical cross-checks, educational use, and benchmarking of more complex neutrino oscillation software, rather than as a replacement for full experimental simulation frameworks. (Here, we consider the DUNE experiment's baseline length in the python implementation but in general we can implement any value of baseline length ). 
\end{abstract}

\begin{keyword}
Neutrino oscillations \sep NeutrinoOsc3Flavor \sep Matter effects \sep CP Violation \sep Cardano change in Variables
\end{keyword}

\end{frontmatter}


\section*{PROGRAM SUMMARY}

\noindent
\textbf{Manuscript Title:} Computational study of three-flavor neutrino oscillations in vacuum and matter with CP-phase sensitivity

\noindent
\textbf{Authors:} Baktiar Wasir Farooq, Bipin Singh Koranga, Ansh Prasad, Imran Khan

\noindent
\textbf{Program Title:} NeutrinoOsc3Flavor

\noindent
\textbf{Programming Language:} Python

\noindent
\textbf{Developer’s repository link: }\url{https://github.com/baktiar238/NeutrinoOsc3Flavor}

\noindent
\textbf{Licensing provisions}: MIT License

\noindent
\textbf{Supplementary material:}
Example input files and sample output data are provided in the source repository.

\noindent
\textbf{Operating System:} 
Platform independent (Linux, Windows, macOS)

\noindent
\textbf{Keywords:} Neutrino oscillation, matter effects, CP violation, numerical simulation

\noindent
\textbf{External Libraries:} NumPy and Matplotlib

\noindent
\textbf{Does the new version supersede the previous version?:} No

\noindent
\textbf{Reasons for the new version:}  
Initial release 

\noindent
\textbf{Nature of Problem:}  
Computation of three-flavor neutrino oscillation probabilities in vacuum and matter. This includes investigating the sensitivity of CP-violating phase and matter density in neutrino oscillation and survival probabilities. The program is designed to solve the Schrödinger evolution equation in the flavor basis for propagation both in vacuum and constant matter density. It serves as a pedagogical and verification-oriented tool for theoretical cross-checks.

\noindent
\textbf{Solution Method:}  
The framework numerically solves the Schrödinger evolution equation by explicitly constructing the effective Hamiltonian within the PMNS formalism. It utilizes numerical diagonalization via the NumPy library to obtain the evolution operator. For internal consistency and validation, it also implements an independent analytical determination of Hamiltonian eigenvalues in matter using the Cardano method.

\noindent
\textbf{Additional comments including restrictions and unusual features:} 
\begin{itemize}
    \item \textbf{Restrictions:} The code currently utilizes a constant matter density approximation (e.g., $\rho=2.7g/cm^{3}$) and neglects neutrino absorption effects.
    \item \textbf{Unusual Features:} It includes a unique independent analytical validation of Hamiltonian eigenvalues via the Cardano method. It also features explicit CP-phase sensitivity studies for both vacuum and matter cases.
    \item Designed as a minimal-dependency implementation relying solely on NumPy and Matplotlib, ensuring seamless execution across Linux, Windows, and macOS.
\end{itemize}
\noindent
\textbf{Running Time:}  
Typically of order milliseconds per probability calculation on a standard desktop CPU.

\section{Introduction}

Neutrino oscillation physics has reached a level of experimental and theoretical precision where numerical reliability and theoretical transparency are as important as raw computational capability. Modern oscillation studies---particularly those involving matter effects and CP-violating phase dependence---are often carried out using large, highly optimized software frameworks developed primarily for experimental simulation and data analysis. While these tools are indispensable for large-scale studies, they typically function as \emph{black boxes} from the perspective of theorists: the underlying numerical procedures, intermediate steps, and even some physics assumptions are deeply embedded in complex software stacks, making independent verification and equation-level inspection difficult.

In this context, there is a clear need for \emph{transparent, minimal, and verification-oriented reference implementations} that allow researchers to directly trace every step of the oscillation calculation---from the construction of the effective Hamiltonian to the evaluation of the evolution operator and final probabilities. Such tools are particularly valuable for sensitivity studies, theoretical cross-checks, pedagogical applications, and benchmarking of more elaborate simulation codes.

In this work, we present \texttt{NeutrinoOsc3Flavor}, a lightweight and dependency-free computational framework for exact three-flavor neutrino oscillation calculations in vacuum and constant-density matter. Rather than optimizing for experimental throughput or large-scale parameter scans, the framework is intentionally designed around \emph{equation-level accessibility}: every component of the oscillation formalism is implemented explicitly and transparently within the flavor-basis Schr\"odinger evolution equation. This design allows users to inspect, modify, and validate each stage of the calculation without reliance on hidden numerical routines or external compiled libraries.

A central design principle of \texttt{NeutrinoOsc3Flavor} is \emph{minimalist dependency and maximal portability}. The entire framework is implemented in pure Python and relies only on NumPy for numerical linear algebra. No external C++ or FORTRAN compilers, specialized numerical libraries, or platform-dependent installations are required. As a result, the code executes natively across Linux, Windows, and macOS systems, enhancing reproducibility and lowering the barrier to independent verification.

Beyond numerical diagonalization, the framework incorporates an \emph{independent analytical determination of the matter-modified Hamiltonian eigenvalues using the Cardano method}. This implementation is not presented merely as an additional computational feature, but as a \emph{benchmarking and verification standard}. By comparing analytical eigenvalues with those obtained via numerical diagonalization, the framework provides an internal consistency check that can be used by other researchers to validate their own oscillation codes and numerical solvers.

The present implementation places particular emphasis on \emph{full CP-violating phase dependence} in both vacuum and matter. Rather than treating the CP phase as a fixed or secondary parameter, \texttt{NeutrinoOsc3Flavor} is specifically structured to explore CP-phase sensitivity as a diagnostic tool for numerical stability and physical correctness. This focus positions the framework as a specialized tool for CP-sensitivity studies and theoretical validation, rather than as a general-purpose experimental simulation package.

\section{Theory}
In this section, we will be discussing about the theoretical formalism of the Neutrino Oscillations in three-flavor in vacuum and in matter. The analytical approach for computing the Neutrino Oscillations in three-flavor oscillations via eigenvalues will also be explicitly discussed for the matter case. We have referred papers \cite{6} - \cite{11}.
\subsection{Neutrino oscillations in vacuum}

Neutrino flavor eigenstates $\ket{\nu_\alpha}$ are related to mass eigenstates $\ket{\nu_i}$
through the PMNS matrix,
\begin{equation} \label{1}
\left(\begin{array}{c}
     \nu_{e} \\
     \nu_{\mu}\\
     \nu_{\tau}
\end{array}\right)=\left(\begin{array}{ccc}
c_{12}c_{13} & s_{12}c_{13} & s_{13}e^{-i\delta}\\
-s_{12}c_{23}-c_{12}s_{23}s_{13}e^{i\delta} & c_{12}c_{23}-s_{12}s_{23}s_{13}e^{i\delta} & s_{23}c_{13}\\
s_{12}s_{23}-c_{12}c_{23}s_{13}e^{i\delta} & -c_{12}s_{23-}s_{12}s_{13}s_{23}e^{i\delta} & c_{23}s_{13}
\end{array}\right)\left(\begin{array}{c}
     \nu_{1} \\
     \nu_{2}\\
     \nu_{3}
\end{array}\right),    
\end{equation}
This equation can be written in the compact form as shown:
\begin{equation} \label{2}
\ket{\nu_\alpha} = \sum_k U_{\alpha k}^* \ket{\nu_k}.
\end{equation}
Here, the $\ket{\nu_{\alpha}}$ represents the Neutrino flavor basis, $U_{\alpha k}^*$ represents the unitary P.M.N.S matrix and the $\ket{\nu_k}$ represents the mass eigenstates. \\
We study the evolution of neutrino flavors as a function of the time (t) or baseline length (L). Solving the Schrödinger equation for neutrino propagation in vacuum in the mass basis, and expressing the result in the flavor basis, one obtains:
\begin{equation} \label{3}
\ket{\nu_\alpha (t)} = \sum_k U_{\alpha k}^* e^{-iE_{k}t} \ket{\nu_k}. 
\end{equation}
where $E_{k}$ represents the energy eigenvalues of the mass eigenstates. To get the Oscillation Probability of the Neutrinos from flavor $\alpha$ to $\beta$ over time (t) can we represented as:
\begin{equation} \label{4}
    P(\nu_{\alpha}\rightarrow\nu_{\beta})=|\bra{\nu_{\beta}\ket{\nu_{\alpha}(t)}}^2 =\sum_{k,j=1} U^*_{\alpha k} U_{\beta k}U_{\alpha j}U^*_{\beta j} e^{i(E_{k}-E_{j})t}
\end{equation}
Here, the $E_{k}$ and $E_{j}$ are the energy of the neutrino mass eigenstates. Using the ultrarelativistic case (equal momentum approximation) and the natural units $c=\hbar=1$ we finally get the expression of probability in terms of mass-square difference and baseline length (L):
\begin{equation} \label{5}
   \Delta E_{kj}= \frac{(m_{k}^2 - m_{j}^2)}{2E} = \frac{\Delta_{kj}}{2E}
\end{equation}
This is the equal momentum approximation and hence, using this the the expression of probability can be expressed as:
\begin{equation} \label{6}
    P(\nu_{\alpha}\rightarrow\nu_{\beta})=\sum_{k,j=1} U^*_{\alpha k} U_{\beta k}U_{\alpha j}U^*_{\beta j} e^{i(\frac{\Delta_{kj}}{2E})L}
\end{equation}
where E is the average energy of the neutrino flux observed after traveling a length of L.\\
On further modifications, we finally get:
\begin{align} \label{7}
P(\nu_{\alpha}\rightarrow\nu_{\beta}) &= \delta_{\alpha \beta} 
- 4 \, \text{Re} \sum_{k,j=1} U^*_{\alpha k} U_{\beta k} U_{\alpha j} U^*_{\beta j} 
  \, \sin^2\left(\frac{\Delta_{kj} L}{4E}\right) \nonumber \\
&\quad + 2 \, \text{Im} \sum_{k,j=1} U^*_{\alpha k} U_{\beta k} U_{\alpha j} U^*_{\beta j} 
  \, \sin\left(\frac{\Delta_{kj} L}{2E}\right)
\end{align}
This is the final formula for getting the oscillation probability of neutrinos, to get the survival probability one just needs to replace the $\beta$ to $\alpha$. One of the most important point to focus in \eqref{7} is the phase associated with the sine terms. The phase is always dimensionless so we replace $1/4$ by 1.27 and for the $1/2$, we just multiply the previous factor by 2. This is the reason we replace these fractions when we build the program for neutrino oscillation calculations in km/GeV units.
\subsection{Matter effects}
The situation becomes more complicated in the matter case. This is because unlike the vacuum case, the exact values of the the mass square differences are not directly known . So, one must solve the effective Hamiltonian in matter ($H_{f}^m$), and obtain its eigenvalues which are the modified mass square differences. In the program, we use the module \texttt{numpy} for calculating the modified PMNS matrix and the eigenvalues or the mass square difference. Also, we have developed an analytical method to accurately calculate the modified mass square difference and we have compared the accuracy of the results with the  computational values. We have referred the works of \cite{4} and \cite{5}  to make the formalism in the matter case. \\
The time evolution of the Neutrino flavor represented by a function of L is represented by:
\begin{equation} \label{8}
   \ket{\nu_{f} (L)}= e^{-i \hat{H}_{f} L} \ket{\nu_f}
\end{equation}
Here, $\nu_f$ represents the neutrino flavor states and $\hat{H}_{f}$
is the effective Hamiltonian in the flavor basis. $e^{-i \hat{H}_{f} L}$ is the evolution operator and is expressed by $S(L)$. The probability amplitude for the oscillation of flavor $\alpha \rightarrow \beta$ is given by:
\begin{equation} \label{9}
    A_{\alpha \rightarrow \beta}=S_{\beta \alpha}(L)=\bra{\nu_{\beta}}S(L) \ket{\nu_{\alpha}}
\end{equation}
We take the same momentum approximation \eqref{5} and solve the Schrödinger equation for the mass eigenstates:
\begin{equation} \label{10}
    \hat{H}_k \ket{\nu_k}= E_{k} \ket{\nu_k}
\end{equation}
and hence, we further solve:
\begin{equation} \label{11}
i \frac{d}{dt} 
\begin{pmatrix} 
    \nu_1 \\ \nu_2 \\ \nu_3 
\end{pmatrix} 
=
\begin{pmatrix} 
    E_1 & 0 & 0 \\ 
    0 & E_2 & 0 \\ 
    0 & 0 & E_3 
\end{pmatrix} 
\begin{pmatrix} 
    \nu_1 \\ \nu_2 \\ \nu_3 
\end{pmatrix}
\end{equation}

and finally after further modifications, in terms of flavor states we get:
\begin{equation} \label{12}
    \hat{H}_f \begin{pmatrix}
    \nu_e \\ \nu_{\mu} \\ \nu_{\tau}
    \end{pmatrix} = U \left[\frac{1}{2E} \left(\begin{matrix}
        0 & 0 & 0 \\
        0 & \Delta_{21} & 0 \\
        0 & 0 & \Delta_{31} \\
    \end{matrix}\right)\right] U^{\dagger} \begin{pmatrix}
    \nu_e \\ \nu_{\mu} \\ \nu_{\tau}
    \end{pmatrix}
\end{equation}
this is what we get for the vacuum case, for the matter case the effective Hamiltonian has a potential term due to the charged current interactions. During charged current interactions, the neutrinos interact with lepton of its same kind and as the density of electrons present in the earth matter is much more than that of muon and tau, so due to the charged current interactions, the $\nu_e$ are only neutrino flavor which are affected the most. As a result, the probability due to the c.c interactions the effective Hamiltonian is modified first, which then changes the probability. The neutral current do not affect the oscillation probability because all the flavors of neutrino gets affected by the same amount and hence, the phase changes by equal amount leading to no such effect.\\
So, the finally the effective Hamiltonian is:
\begin{equation} \label{13}
    \hat{H}_f^m = U \left[\frac{1}{2E} \left(\begin{matrix}
        0 & 0 & 0 \\
        0 & \Delta_{21} & 0 \\
        0 & 0 & \Delta_{31} \\
    \end{matrix}\right)\right] U^{\dagger} + \left(\begin{matrix}
        V_{cc} & 0 & 0 \\
        0 & 0 & 0 \\
        0 & 0 & 0 \\ 
    \end{matrix}\right)
\end{equation}
in \eqref{13}, the $V_cc$ is the potential introduced due to the charged current interaction and is given by $V_{cc}=\sqrt{2} G_F n_{e}$ where $G_F$ is the Fermi's Constant and $n_e$ is the number density of electron in the earth matter (we consider the crust).

\subsubsection{Diagonalizing the effective Hamiltonian using \texttt{numpy}:}
Unlike the previous case, where we already knew the mass squared difference, in this case we don't have such approach. So, we need to diagonalize the  \eqref{13}. The modified PMNS matrix in the matter is represented by $W$. So, we can express the flavor basis in matter as:
\[ \ket{\nu_\alpha} = \sum_k W_{\alpha k}^* \ket{\nu_k^m}\]
and, we can write the Schrodinger Equation as:
\[\hat{H}_f^m W = W {\Lambda} \]
$\ket{\nu_k^m}$ is the mass eigenstates in matter and ${\Lambda}$ is the diagonalized eigenvalue matrix, expressed as:
\begin{equation} \label{14}
    {\Lambda} = \begin{pmatrix}
        \lambda_1 & 0 & 0 \\
        0 & \lambda_2 & 0 \\
        0 & 0 & \lambda_3 \\
    \end{pmatrix}
\end{equation}
So, finally the effective Hamiltonian can be written as:
\begin{equation} \label{15}
    \hat{H}_f^m = W\Lambda W^{\dagger}
\end{equation}
Hence, in analogy to equation \eqref{8},\eqref{9} and \eqref{15} $S_{\beta\alpha}(L)$ can be written as:
\[S(L)=e^{-iW\lambda W^\dagger L}\]
Further, by using Taylor series, we finally get S(L) as:
\begin{equation} \label{16}
    S(L)=W\left[\begin{matrix}
        e^{-i \lambda_1L} & 0 & 0 \\
        0 & e^{-i \lambda_2L} & 0 \\
        0 & 0 & e^{-i \lambda_3L}
    \end{matrix}\right] W^\dagger
\end{equation}
Hence, the probability of the oscillation of flavor by squaring the probability amplitude in \eqref{9} can be given by:
\begin{equation} \label{17}
    P_{\beta \rightarrow \alpha} = \left|\sum_k^3 W_{\beta k}|S(L)|W^*_{\alpha k}\right|^2
\end{equation}

here, due to the orthonormal condition: \[\braket{\nu_j^m}{\nu_k^m}=\delta_{jk}\]  there is no mass eigenstate in \eqref{17} as we consider k=j.\\
Under the project - \texttt{NeutrinoOsc3Flavor}, we have developed a python program using the module \texttt{numpy} to directly determine the eigenvalues and eigenvectors i.e, $\lambda$ and $W$, respectively.\\
\[ W,\lambda ={np.linalg.eigh(H_f^m)}\]
This python command will directly solve the effective Hamiltonian and will provide the Oscillation and Survival Probabilities of neutrinos. The details of the programming will be covered in the next section.
\subsection{Analytical Approach to get the eigenvalues of \texorpdfstring{$H_f^m$}{H	extasciimacron{}f	extasciimacron{}m}:} \label{analytical}
We have also formulated an analytical formula for getting the eigenvalue and, compared the results with the computationally determined values. The analytical values were very precise and accurate to those of the computationally determined values. This section focuses on the analytical formalism of the eigenvalues for the effective Hamiltonian matrix.\\
The equation \eqref{13} in compact form can be written as - 
\begin{equation} \label{18}
    H^{\text{m}}_{ij}
= \sum_{k=1}^{3} U_{ik}\, m_i\, U^{*}_{jk} 
+ \underbrace{(V_{\text{C.C}})_{ij}}_{\equiv V\, \delta_{i1}\delta_{j1}} \qquad \text{i, j} \in \{1, 2, 3\}
\end{equation}

Here, $m_2 = \frac{\Delta_{21}}{2E}, \qquad 
m_3 = \frac{\Delta_{31}}{2E}, \, \And \qquad
m_1 = 0$.\\
Now, evaluating this further to form the final effective Hamiltonian Matrix:
\begin{equation} \label{19}
    H^{\text{m}}_f =
\begin{bmatrix}
H^{\text{m}}_{11} & H^{\text{m}}_{12} & H^{\text{m}}_{13} \\
H^{\text{m}*}_{12} & H^{\text{m}}_{22} & H^{\text{m}}_{23} \\
H^{\text{m}*}_{13} & H^{\text{m}*}_{23} & H^{\text{m}}_{33}
\end{bmatrix}
\end{equation}

To obtain the eigenvalues, one must resolve the characteristic polynomial: 
\begin{equation} \label{21}
    \left| H_f^{\text{m}} - \lambda I \right| = 0
\end{equation}
so the final characteristic equation looks like:
\begin{equation} \label{22}
   \lambda^3 + p\lambda^2 + q\lambda + r = 0 
\end{equation} 
Solving the equation \eqref{21}, the final equation turns out to be:
\begin{equation} \label{23}
    \lambda^3  -T_{r}(H_{f}^{m}) \lambda^2 + \frac{1}{2}
\left[T_{r}(H)^2 - T_{r}(H)^2)\right]\lambda  -\det(H_f) = 0 
\end{equation}
Now, as we use the \textbf{Cardano change of variable}, we get the formula of the eigenvalues for solving the effective Hamiltonian ($H_f^m$). The equation turns out to be:
\begin{equation}\label{24}
\begin{aligned}
\lambda_j
&=
\left(
-\frac{b}{2}
+
\sqrt{
\left(\frac{b}{2}\right)^2
+
\left(\frac{a}{3}\right)^3
}
\right)^{1/3}
\omega^{j}
\\[6pt]
&\quad+
\left(
-\frac{b}{2}
-
\sqrt{
\left(\frac{b}{2}\right)^2
+
\left(\frac{a}{3}\right)^3
}
\right)^{1/3}
\omega^{-j}
\;-\;
\frac{\mathrm{Tr}\!\left(H_f^{\text{matter}}\right)}{3},
\\
&\hspace{5cm} j = 0,1,2 .
\end{aligned}
\end{equation}
In this equation \eqref{24}, $\omega=e^{i\big(\frac{2\pi}{3}\big)}$ $a=q - \frac{p^{2}}{3},$ and $b=\frac{2p^{3}}{27} - \frac{2qp}{3} + r$. The values of p,q and r can be determined by comparing equation \eqref{22} and \eqref{23}. 
\section{Code implementation in NeutrinoOsc3Flavor }

Our program - \texttt{NeutrinoOsc3Flavor}, is implemented entirely in Python using \texttt{numPy} for numerical linear algebra and \texttt{matplotlib} for visualization.
The effective Hamiltonian is constructed in the flavor basis and diagonalized numerically to obtain the evolution operator. We referred the works of: \cite{12} - \cite{14}. 



\subsection{Three Flavor Neutrinos in Vacuum:}
In this section, we will be focusing on the Three flavor case of Neutrino Oscillations in Vacuum.\\
The program \texttt{NeutrinoOsc3Flavor} focuses on the variation of neutrino oscillation probabilities with respect to changes in the CP-violating phase. The CP phase values considered in this study are
\[
0^\circ,\; 45^\circ,\; 90^\circ,\; 135^\circ,\; 180^\circ,\; \text{and } 217^\circ .
\]
\begin{lstlisting}
CP_phase = np.deg2rad([0,45,90,180,217])
for CP in CP_phase:
    def U_matrix(CP,t12,t23,t13):
        s12,s23,s13,sCP=np.sin(t12),np.sin(t23),np.sin(t13),np.sin(CP)
        c12,c23,c13,cCP=np.cos(t12),np.cos(t23),np.cos(t13),np.cos(CP)
        U=np.array([
            [c12*c13,   s12*c13,   s13*np.exp(-1j*CP)],
            [-s12*c23-c12*s23*s13*np.exp(1j*CP),                          c12*c23-s12*s23*s13*np.exp(1j*CP),   s23*c13],
            [s12*s23-c12*c23*s13*np.exp(1j*CP),                           c12*s23-s12*c23*s13*np.exp(1j*CP),   c23*c13]
            ],dtype=complex)
        return U
\end{lstlisting}
The above code will hence run for the allotted values of CP Phases. For the Oscillation Probabilities, we used the analytical formulas \eqref{7}:
\begin{lstlisting}
    U=U_matrix(CP,t12,t23,t13)
    
    # Real parts
    U1_real = (U[0][1].conj()*U[1][1]*U[0][0]*U[1][0].conj()).real 
    U2_real = (U[0][2].conj()*U[1][2]*U[0][1]*U[1][1].conj()).real
    U3_real = (U[0][2].conj()*U[1][2]*U[0][0]*U[1][0].conj()).real

    # Imaginary parts
    U1_img = (U[0][1].conj()*U[1][1]*U[0][0]*U[1][0].conj()).imag
    U2_img = (U[0][2].conj()*U[1][2]*U[0][1]*U[1][1].conj()).imag
    U3_img = (U[0][2].conj()*U[1][2]*U[0][0]*U[1][0].conj()).imag

#FOR THE OSCILLATION PROBABILITY OF ELECTRON TO MUON NEUTRINOS:
    Osc_Probability_mu = (
        -4 * (U1_real * (np.sin(1.27 * delta12 * LbyE))**2 +
              U2_real * (np.sin(1.27 * delta23 * LbyE))**2 +
              U3_real * (np.sin(1.27 * delta13 * LbyE))**2)
        + 2 * (U1_img * np.sin(2.54 * delta12 * LbyE) +
               U2_img * np.sin(2.54 * delta23 * LbyE) +
               U3_img * np.sin(2.54 * delta13 * LbyE))
    )

\end{lstlisting}
In this section we have used the codes for the oscillation from the $\nu_e \rightarrow \nu_{\mu}$. This program is very flexible as one can take for any flux of neutrino flavors like $\nu_e,\nu_{\mu}, \nu_{\tau}$. One must change 0 to 1 for $\nu_{\mu}$ in a way:\\

\begin{lstlisting}
     # Real parts
    U1_real = (U[1][1].conj()*U[1][1]*U[1][0]*U[1][0].conj()).real 
    U2_real = (U[1][2].conj()*U[1][2]*U[1][1]*U[1][1].conj()).real
    U3_real = (U[1][2].conj()*U[1][2]*U[1][0]*U[1][0].conj()).real

    # Imaginary parts
    U1_img = (U[1][1].conj()*U[1][1]*U[1][0]*U[1][0].conj()).imag
    U2_img = (U[1][2].conj()*U[1][2]*U[1][1]*U[1][1].conj()).imag
    U3_img = (U[1][2].conj()*U[1][2]*U[1][0]*U[1][0].conj()).imag

\end{lstlisting}
This modification serves the survival probability of muon neutrinos ($\nu_{\mu} \rightarrow \nu_{\mu}$).\\
Note: $0 \equiv \nu_e$ , $1 \equiv \nu_{\mu}$ and $2 \equiv \nu_{\tau}$. These logics can be used to modify the program for any neutrino flavors.

\subsection{Three Flavor Neutrinos in Matter:}
This section contains the method we developed to get the eigenvalues for the matter dependent Hamiltonian. We calculated the eigenvalues via \texttt{numpy} as well as the developed analytical formula shown in\eqref{24}.

\subsubsection{Eigenvalues using numpy:} \label{used codes}

\begin{lstlisting}
    #Main loop over CP phases
for CP in deltaCP_list:
    U = U_pmns(t12, t13, t23, CP)
    P_ue_vals, P_uu_vals, P_ut_vals = [], [], []

    for Ei in E:
        E_eV = Ei * E_GeV_2_eV
        H_f_vac = U @ np.diag([0, dm21/(2*E_eV), dm31/(2*E_eV)]) @ U.conjugate().T
        H_f_matter = H_f_vac + np.diag([V, 0, 0])
        
#Computational results:
        evals, evecs = np.linalg.eigh(H_f_matter)
        W, lambdas = evecs, evals
\end{lstlisting}
On printing the values of lambda, we get the eigenvalues of Hamiltonian matrix. This approach directly solves the Hamiltonian and hence, we further use to develop the formulas for plotting the oscillation probabilities in the matter. 

\subsubsection{Eigenvalues using analytical method:}
\begin{lstlisting}
            p = -np.trace(H_f_matter)
        q = 0.5 * (np.trace(H_f_matter)**2 - np.trace(H_f_matter@H_f_matter))
        r = -np.linalg.det(H_f_matter)

# Depressed cubic: x = lamda + p/3  to  x^3 + a x + b = 0
        a = q - (p**2)/3
        b = (2*p**3)/27 - (p*q)/3 + r

# Discriminant terms
        delta = (b/2)**2 + (a/3)**3

# Cube roots (principal complex cube roots)
        C_plus  = (-b/2 + np.sqrt(delta))**(1/3)
        C_minus = (-b/2 - np.sqrt(delta))**(1/3)

# Cube roots of unity
        omega = np.exp(1j * 2*np.pi/3)

# Three roots (lamda_j)
        lamda = []
        for k in range(3):
            xj = C_plus * omega**(k) + C_minus * omega**(-k)
            lamdaj = xj - p/3  # undo shift (since p = -Tr(H))
            lamda.append(lamdaj.real)

\end{lstlisting}

This approach is introduced in \ref{analytical}.
\vspace{2cm}
 \subsubsection{Plots for the Probabilities in matter:}
 \begin{lstlisting}
      # probabilities (note: indices are S[beta, alpha])
        A_ue = S[0, 1]
        A_uu = S[1, 1]
        A_ut = S[2, 1]

        P_ue_vals.append(np.abs(A_ue)**2)
        P_uu_vals.append(np.abs(A_uu)**2)
        P_ut_vals.append(np.abs(A_ut)**2)
        
# Probability expression:
    P_ue_vals = np.array(P_ue_vals)
    P_uu_vals = np.array(P_uu_vals)
    P_ut_vals = np.array(P_ut_vals)

 \end{lstlisting}
In the above program, we have used the eigenvalues and eigenvectors in \ref{used codes} to determine the oscillation probability of the neutrino flavors.

We have verified our results by plotting the Oscillation Probabilities with respect to the energy and compared it with \cite{2} for the vacuum case and for the matter case, we compared the oscillation probabilities with \cite{3}
\begin{figure}[H]
  \centering
  \begin{subfigure}[t]{0.48\textwidth}
    \centering
    \includegraphics[height=5cm, keepaspectratio]{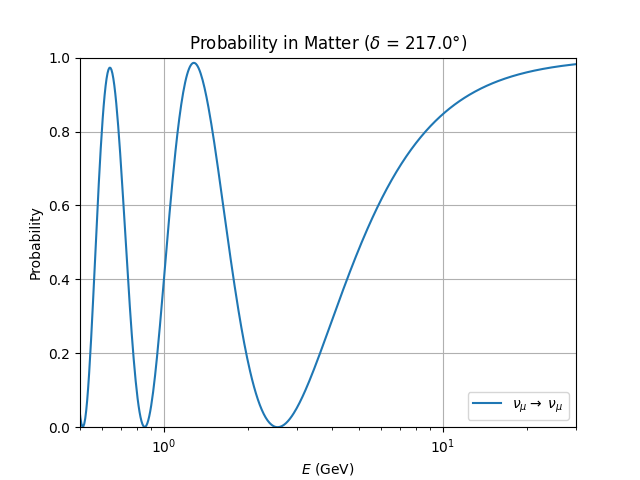}
    \label{fig:left}
  \end{subfigure}
  \hfill
  \begin{subfigure}[t]{0.50\textwidth}
    \centering
    \includegraphics[height=5cm, keepaspectratio]{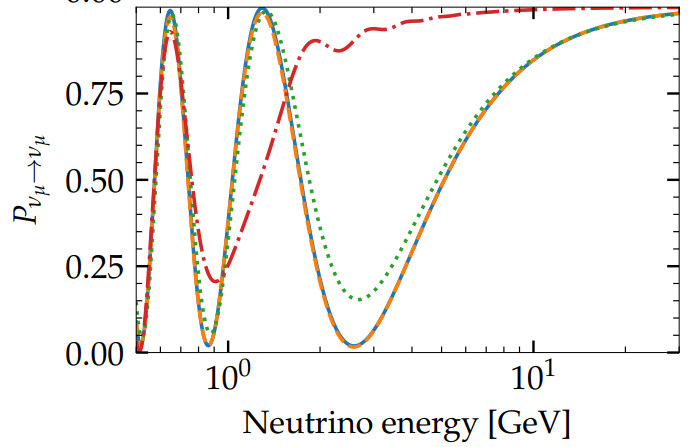}
    \label{fig:right}
  \end{subfigure}
  \caption{Comparison of muon neutrino survival probabilities in matter.The left panel shows the result obtained using \texttt{NeutrinoOsc3Flavor}, while the right panel shows the corresponding result obtained using \texttt{NuOscProbExact}~\cite{3}. The orange dashed curve represents the survival probability in matter.}
  \label{fig:twofigures}
\end{figure}

\section{Results}

\subsection{\texorpdfstring{Comparison of the eigenvalues of the $H_f^m$ matrix}
{Comparison of the eigenvalues of the Hfm matrix}}
\begin{lstlisting}
   lamdas (analytical)= [np.float64(8.721563027439424e-16), np.float64(4.023643281436783e-14),np.float64(1.1520807754955487e-13)]
lamdas (computational)= [8.72156303e-16 4.02364328e-14 1.15208078e-13]
\end{lstlisting}
Here, we can observe that the analytical solution provides an independent cross-check of the numerical diagonalization.
\subsection{Variation with CP Phase for the Vacuum case:}
\begin{figure}[H]
    \centering
    
    \begin{subfigure}[b]{0.45\textwidth}
        \centering
        \includegraphics[width=\textwidth]{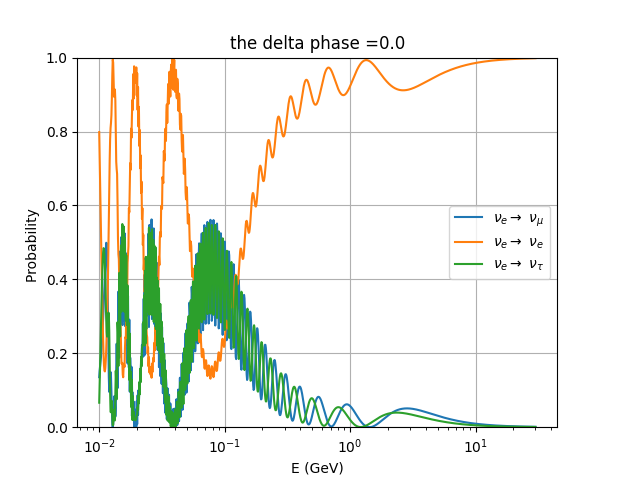}
        \caption{Solar Neutrino Oscillation Probability when $\delta_{CP}=0^\circ$}
       
    \end{subfigure}
    \hfill
    \begin{subfigure}[b]{0.45\textwidth}
        \centering
        \includegraphics[width=\textwidth]{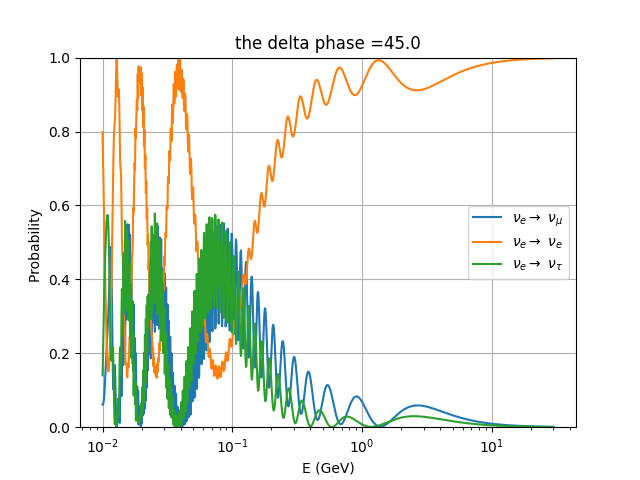}
        \caption{Solar Neutrino Oscillation Probability when $\delta_{CP}=45^\circ$}
        
    \end{subfigure}
    
    \vspace{0.5cm} 
    
    \begin{subfigure}[b]{0.45\textwidth}
        \centering
        \includegraphics[width=\textwidth]{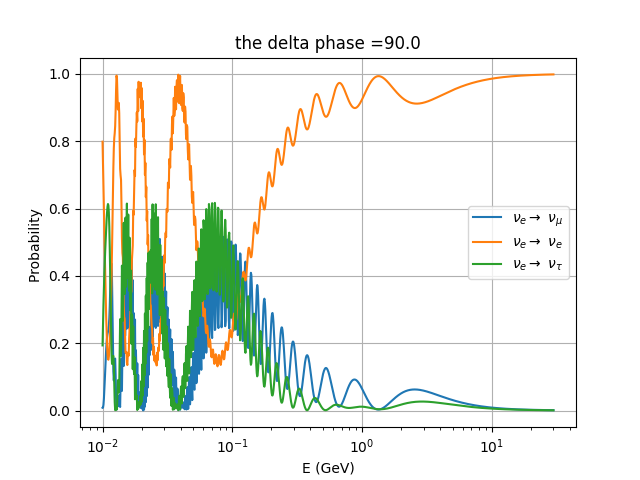}
        \caption{Solar Neutrino Oscillation Probability when $\delta_{CP}=90^\circ$}
       
    \end{subfigure}
    \hfill
    \begin{subfigure}[b]{0.45\textwidth}
        \centering
        \includegraphics[width=\textwidth]{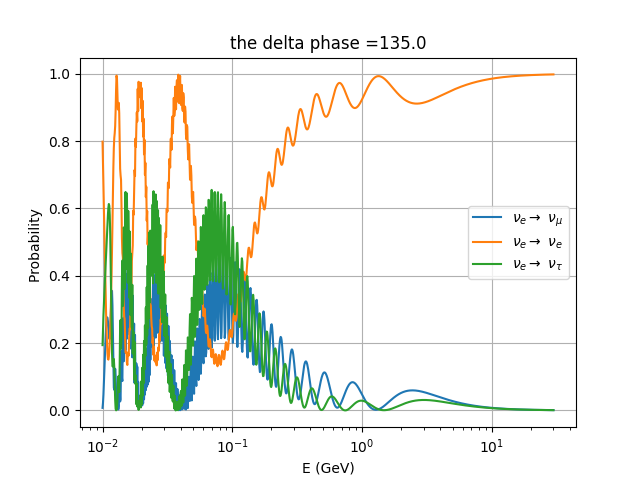}
        \caption{Solar Neutrino Oscillation Probability when $\delta_{CP}=135^\circ$}
       
    \end{subfigure}
    
    \vspace{0.5cm} 
    
    \begin{subfigure}[b]{0.45\textwidth}
        \centering
        \includegraphics[width=\textwidth]{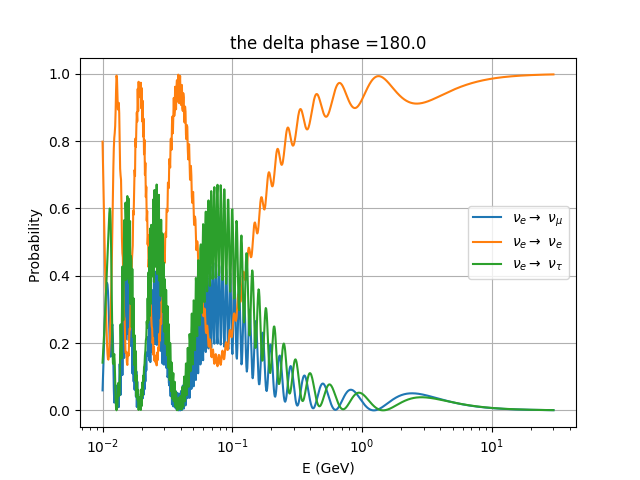}
        \caption{Solar Neutrino Oscillation Probability when $\delta_{CP}=180^\circ$}
        
    \end{subfigure}
    \hfill
    \begin{subfigure}[b]{0.45\textwidth}
        \centering
        \includegraphics[width=\textwidth]{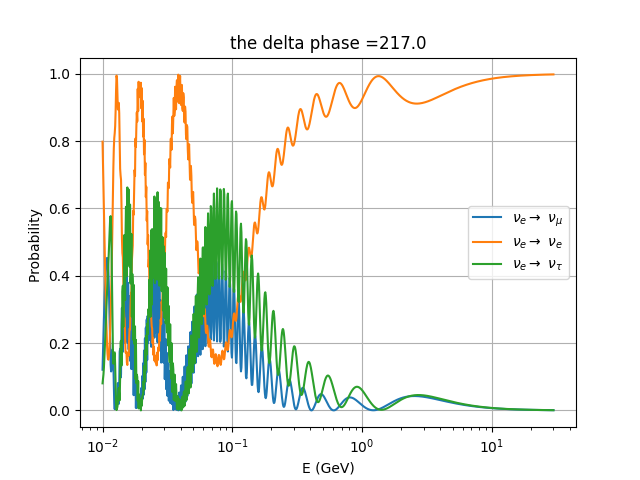}
        \caption{Solar Neutrino Oscillation Probability when $\delta_{CP}=217^\circ$}
       
    \end{subfigure}
    
    \caption{In vacuum, the oscillation probability will have changes but not the survival probability}
\end{figure}

\subsection{Variation with CP Phase for the Matter case:}
\begin{figure}[H]
    \centering
    
    \begin{subfigure}[b]{0.45\textwidth}
        \centering
        \includegraphics[width=\textwidth]{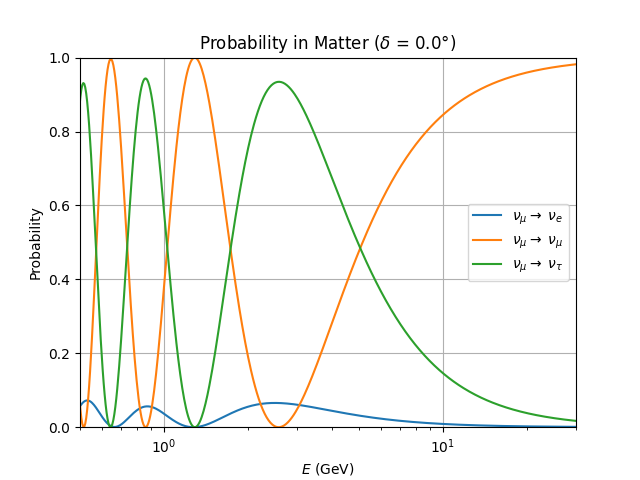}
        \caption{Atmospheric Neutrino Oscillation Probability when $\delta_{CP}=0^\circ$}
       
    \end{subfigure}
    \hfill
    \begin{subfigure}[b]{0.45\textwidth}
        \centering
        \includegraphics[width=\textwidth]{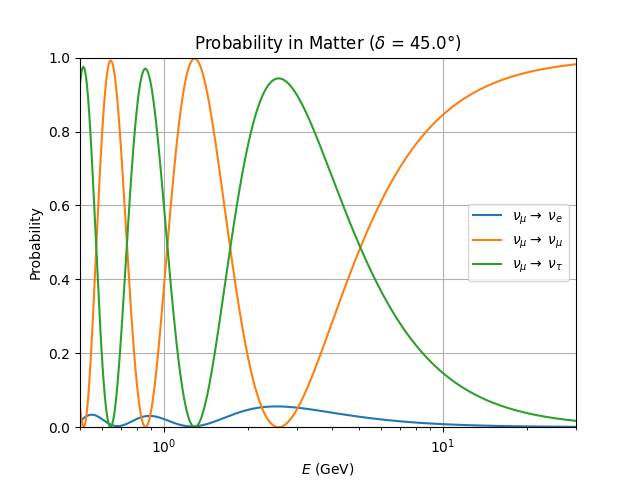}
         \caption{Atmospheric Neutrino Oscillation Probability when $\delta_{CP}=45^\circ$}
        
    \end{subfigure}
    
    \vspace{0.5cm} 
    
    \begin{subfigure}[b]{0.45\textwidth}
        \centering
        \includegraphics[width=\textwidth]{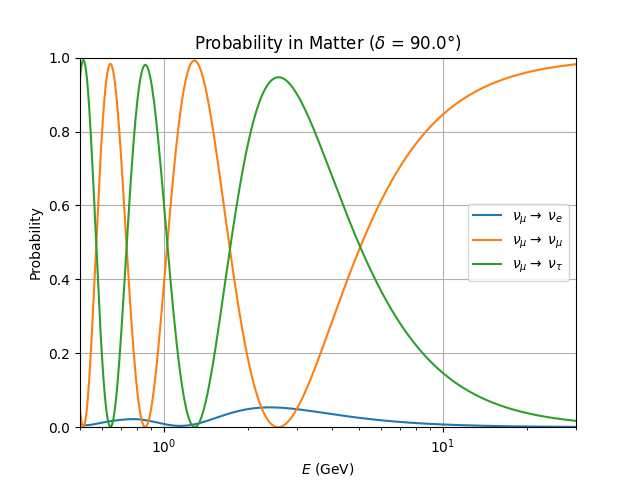}
         \caption{Atmospheric Neutrino Oscillation Probability when $\delta_{CP}=90^\circ$}
        
    \end{subfigure}
    \hfill
    \begin{subfigure}[b]{0.45\textwidth}
        \centering
        \includegraphics[width=\textwidth]{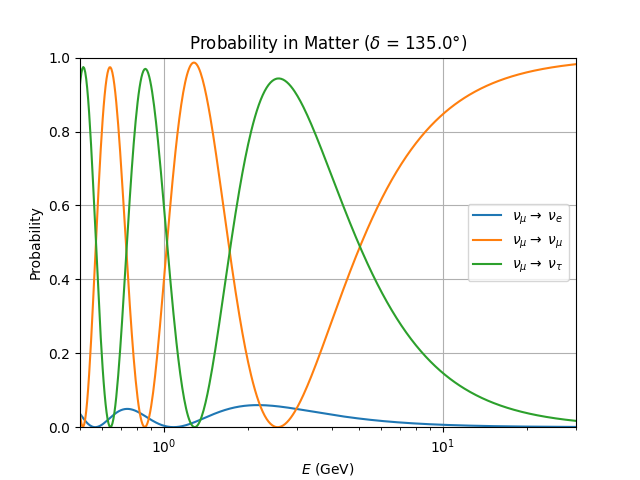}
        \caption{Atmospheric Neutrino Oscillation Probability when $\delta_{CP}=135^\circ$}
        
    \end{subfigure}
    
    \vspace{0.5cm} 
    
    \begin{subfigure}[b]{0.45\textwidth}
        \centering
        \includegraphics[width=\textwidth]{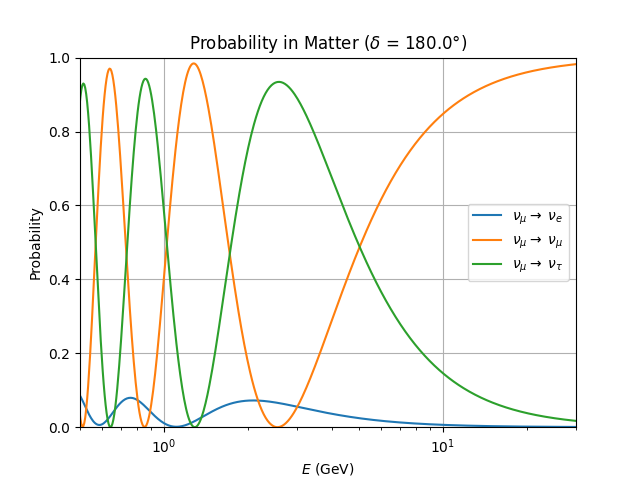}
         \caption{Atmospheric Neutrino Oscillation Probability when $\delta_{CP}=180^\circ$}
        
    \end{subfigure}
    \hfill
    \begin{subfigure}[b]{0.45\textwidth}
        \centering
        \includegraphics[width=\textwidth]{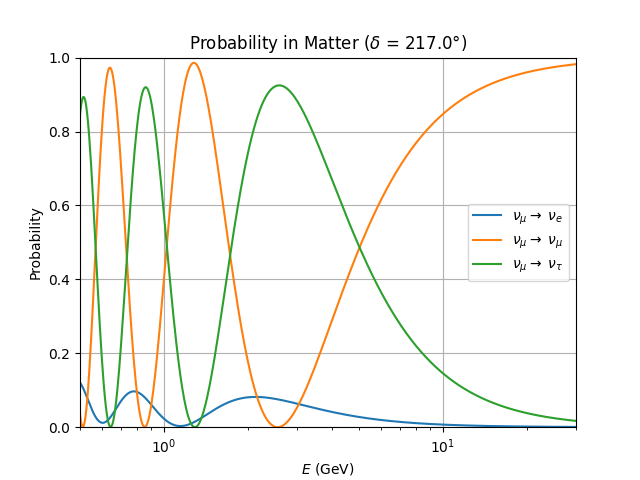}
         \caption{Atmospheric Neutrino Oscillation Probability when $\delta_{CP}=217^\circ$}
        
    \end{subfigure}
    
    \caption{In matter, the oscillation probabilities will have changes and the survival probability will also show slight changes due to matter effects.}
    
\end{figure}

\section{Conclusions}

We have presented \texttt{NeutrinoOsc3Flavor}, a transparent and verification-oriented computational framework for exact three-flavor neutrino oscillation studies in vacuum and constant-density matter. The framework is deliberately designed to prioritize \emph{how oscillation calculations are performed}, rather than merely reproducing known results. By implementing the full Schr\"odinger evolution equation explicitly in the flavor basis, the code provides equation-level access to every step of the physics, making it particularly well suited for theoretical validation and pedagogical use.

A defining feature of \texttt{NeutrinoOsc3Flavor} is its \emph{minimalist software design}. The framework requires no external compilers, specialized numerical libraries, or complex installation procedures, relying solely on NumPy for linear algebra operations. This minimal dependency structure ensures cross-platform portability and reproducibility, and distinguishes the code from large-scale oscillation toolkits whose complexity can obscure numerical assumptions and intermediate results.

The inclusion of an \emph{independent analytical eigenvalue solver based on the Cardano method} plays a central role in the framework’s philosophy. Rather than serving as an auxiliary option, the analytical solution functions as a \emph{benchmarking standard} against which numerical diagonalization can be verified. The excellent agreement observed between analytical and numerical eigenvalues provides a robust internal consistency check and demonstrates the numerical stability of the evolution operator used to compute oscillation probabilities. This feature makes \texttt{NeutrinoOsc3Flavor} particularly valuable as a reference implementation for testing and validating more complex oscillation codes.
The framework’s explicit treatment of \emph{full CP-violating phase dependence} further enhances its role as a specialized verification and sensitivity-analysis tool. By allowing the CP phase to be varied freely and systematically in both vacuum and matter, the code enables detailed studies of CP-induced effects and serves as a diagnostic probe for numerical correctness in three-flavor oscillation calculations. This focus distinguishes the framework from general-purpose simulation tools and positions it as a targeted resource for CP-sensitivity studies.
Overall, \texttt{NeutrinoOsc3Flavor} is best understood not as a replacement for comprehensive experimental simulation frameworks, but as a \emph{transparent, lightweight reference platform} for exact verification, benchmarking, and theoretical cross-checks in neutrino oscillation physics. Its emphasis on analytical traceability, numerical validation, and reproducibility makes it a useful complement to existing large-scale tools and a reliable foundation for future methodological and educational work.

\end{document}